\documentclass[prd,preprintnumbers,nofootinbib,showpacs,12pt]{revtex4-1}

\usepackage{bm}
\usepackage{amsmath,amsbsy,graphicx,amsfonts,mathrsfs,amssymb}
\usepackage{latexsym}
\usepackage{graphicx,epsfig}
\usepackage{pdfsync}
\usepackage{float}
\usepackage{color}
\usepackage[hyperindex,CJKbookmarks=true]{hyperref}
\usepackage{calc}

\newcommand{\f}{\begin{equation}}
\newcommand{\ff}{\end{equation}}
\newcommand{\fa}{\begin{eqnarray}}
\newcommand{\ffa}{\end{eqnarray}}

\begin{document}

\title{Holographic incoherent transport in Einstein-Maxwell-dilaton Gravity}
\author{Zhenhua Zhou $^{1}$}
\email{zhouzh@ihep.ac.cn}
\author{Yi Ling $^{1,3}$}
\email{lingy@ihep.ac.cn}
\author{Jian-Pin Wu $^{2,3}$}
\email{jianpinwu@mail.bnu.edu.cn} \affiliation{$^1$ Institute of
High Energy Physics, Chinese
Academy of Sciences, Beijing 100049, China\\
$^2$ Institute of Gravitation and Cosmology, Department of
Physics, School of Mathematics and Physics, Bohai University,
Jinzhou 121013, China \\ $^3$ State Key Laboratory of Theoretical
Physics, Institute of Theoretical Physics, Chinese Academy of
Sciences, Beijing 100190, China}
\begin{abstract}

Recent progress in the holographic approach makes it more transparent
that each conductivity can be decomposed into the coherent
contribution due to momentum relaxation and the incoherent
contribution due to intrinsic current relaxation. In this paper we
investigate this decomposition in the framework of
Einstein-Maxwell-dilaton theory. We derive the perturbation
equations, which are decoupled for a large class of background
solutions, and then obtain the analytic results of conductivity
with the slow momentum relaxation in low frequency approximation,
which is consistent with the known results from memory matrix techniques.

\end{abstract}
\maketitle

\section{Introduction}

Many strongly interacting many-body systems do not have
quasiparticle excitations, for instance, the strange metal phase
of cuprate superconductors
\cite{Marel:2003,Cooper:2009,Sachdev:2011cs}. It is still
challenging for condensed matter physicists to understand the
transport properties of strongly coupled systems without
quasiparticles at finite temperature. In recent years, holographic
duality provides a new approach for exploring these properties in the
strongly correlated
system\cite{Hartnoll:2009sz,McGreevy:2009xe,Sachdev:2011wg}.
The landmark achievement includes the universal bounds on diffusion rates of momentum, charge, and energy
proposed in \cite{bound1,Hartnoll:2014lpa}. \footnote{Some extended studies have also been explored in \cite{bound2,bound3,bound4}.}

To handle the transport properties in a strongly
interacting system without quasiparticles, a traditional tool is
the memory matrix
framework
\cite{Zwanzig:1961,Mori:1965,Forster:1975,Hartnoll:2007ih,Hartnoll:2014gba,Patel:2014jfa,Jung:2007,Lucas:20151,Lucas:20156},
in which a perturbative expansion with small momentum relaxation
rate $\Gamma$ can be performed. In particular, at leading order of
expansion it can give rise to the universal Drude behavior of
conductivity. Recently, the research on incoherent transport
from the perspective of holography has been developed
\cite{Lucas:20156,G:20141,Hartnoll:2014lpa,Davison:201411,Ling:201412,Davison:2015bea,Davison:2015taa,Ling:2015exa,Donos:201511,Lucas:201511}.
These results can be extended by incorporating the laws of
relativistic conformal hydrodynamics at the phenomenological level(see
\cite{Hartnoll:2007ih} or Eq.(1.3) in \cite{Davison:2015bea}), in
which the conductivity can be decomposed into coherent
contribution due to the momentum relaxation and incoherent
contribution due to intrinsic current relaxation. Since the
incoherent conductivity is decoupled from the momentum
\cite{Hartnoll:2014lpa}, it is supposed to capture the intrinsic
behavior of the electric transport, and then to be a completely
intrinsic characteristic of the system at low energy physics.
Therefore, it is very crucial to understand its universal features
in more  generic circumstances.

For a system with no momentum dissipation, one can define an
incoherent current $J^{inc}$, a particular combination of the
charge and heat currents, which has no overlap with the momentum.
Its corresponding conductivity $\sigma_{J^{inc}J^{inc}}$ (or
$\sigma_{inc}$ in shorthand) is finite and may capture some
universal transport properties. Its direct current (dc) conductivity turns
out to be universal and can be obtained by the hydrodynamic method
for a class of holographic models as\cite{Hartnoll:2007ih,Davison:2015taa}
\begin{alignat}{1}\label{sigmainc}
\sigma_{inc}\equiv\sigma_Q=Z_+\Big(\frac{sT}{\epsilon+P}\Big)^2\,.
\end{alignat}

However, for a system with momentum dissipation, it usually becomes
hard to obtain such incoherent current. Nevertheless, it is shown
in a recent paper \cite{Davison:2015bea} that it may be extracted
out in some specific holographic models, for instance, in the
Einstein-Maxwell (EM) theory with axions. The key observation in this
framework is that the linearized perturbation equations can
be decoupled by redefining perturbation variables such that we can
construct a new orthogonal basis $J_{\pm}$ for the currents
instead of the usual charge current $J$ and heat current $J^Q$. As
a result the corresponding response matrix is diagonal, namely,
$\chi_{J^+J^-}=0$. More importantly, it can be found that these
independent currents go back to $J^{inc}$ and momentum,
respectively, when the momentum dissipation vanishes.  At slow
momentum dissipation, they are responsible for the incoherent and
coherent contribution of the low frequency conductivity,
respectively.

In this paper, we intend to demonstrate that the incoherent part
of conductivity can be extracted out for a large class of
Einstein-Maxwell-dilaton (EMD) gravity theories with momentum
dissipation by decoupling the linearized perturbation equations.
Our work further confirms the universality of incoherent transport
in the  holographic approach. In addition, we systematically derive the
low frequency behavior of conductivity at slow momentum
dissipation for a large class of holographic models, including the
Gubser-Rocha EMD model with vanishing ground state entropy
\cite{Gubser:2009qt}.

Our paper is organized as follows. In Sec.\ref{EMDa}, we
derive the decoupling equations for a class of EMD-axion theories
and define the independent currents. In
Sec.\ref{deG}, we simplify the analysis of the conductivity by
relating it  with the diagonal response matrix of the independent
currents. And then the analytic expression of the low frequency
conductivities in the presence of slow momentum dissipation is
obtained in Sec.\ref{Lfc}. The conclusions and discussions are
present in Sec.\ref{con}.

\section{EMD-axion model}\label{EMDa}
EMD theories have been widely studied in the context of
holographic gravity (for instance, see
\cite{Ling:2013nxa,Zhou:2015dha} and the references therein).
Here we are interested in a class of EMD gravity theories whose
action can be written as a general form
\begin{eqnarray}
S=\int d^4x \sqrt{-g}\left(R-\frac{Z(\varphi)}{4}F^2-\beta(\varphi)\sum_{I=x,y}(\partial\phi_I)^2-\frac{1}{2}(\partial\varphi)^2+V(\varphi)\right)
\,,\label{AcEMD}
\end{eqnarray}
where the coupling terms $Z(\varphi)$, $\beta(\varphi)$ and the potential $V(\varphi)$ are general functions of dilaton field $\varphi$,
which allows us to study a large class of EMD models.
The momentum dissipation is introduced by setting the axions $\phi_x=mx,~\phi_y=my$, as orginally proposed in \cite{Andrade:2013gsa}.
$m$ is an arbitrary real number,  characterizing the strength of momentum dissipation.

We consider the following ansatz for the background, which is supposed to be a black brane with asymptotically anti¨Cde Sitter structure:
\begin{eqnarray}
&&ds^2=-r^2f(r)dt^2+\frac{dr^2}{f(r)r^2}+r^2g(r)(dx^2+dy^2)\,,\label{anmg}\\
&&A=A_t(r)dt\,,~~~~~~\varphi=\varphi(r)
\,.\label{bgAt}
\end{eqnarray}
If the position of the horizon is set at $r_0$ by $f(r_0)=0$, then
the bulk geometry described by the above metric is dual to a CFT
with charge density $q$, entropy density $s$, and  temperature $T$
that are separately given by
\begin{eqnarray}
q=Zr^2g A_t^\prime(r)\,,~~~~~s=4\pi r_0^2g(r_0)\,,~~~~~T=r^2_0f^\prime(r_0)/(4\pi)\,.
\label{cftq}\end{eqnarray}

Now we intend to investigate the transport properties of
conductivity in the dual field theory. To do so we turn on the
following perturbations around the background solution in
(\ref{anmg}) and (\ref{bgAt}):
\begin{eqnarray}
&&
A_x=a_x(r)e^{-i\omega t}\,,\qquad \delta \phi_x:=\chi_x(r)e^{-i\omega t}\label{opur1}
\nonumber
\\
&&
\delta g_{tx}:=r^2g(r)h_{tx}(r)e^{-i\omega t}\,,\qquad \delta g_{rx}:=r^2g(r)h_{rx}(r)e^{-i\omega t}\label{opur2}\,.
\end{eqnarray}
Then, we can obtain three independent linearized perturbation equations, which read as
\begin{eqnarray}
\label{eome1}
&&
2 m^2 \beta h_{tx}-\frac{f}{g}(g^2r^4\psi)^\prime+i\omega2m\beta \chi_x=\frac{qf}{g} a_x^\prime\,,
\
\\
\label{eome2}
&&
2 m^2 \beta h_{rx}+\frac{i\omega g\psi}{f}-2m\beta \chi_x^\prime=-\frac{i\omega q}{r^4fg}a_x\,,
\
\\
&&
(Zr^2fa_x^\prime)^\prime+\frac{\omega^2 Z}{r^2f}a_x+q\psi=0\,.
\end{eqnarray}
where $\psi\equiv h_{tx}^\prime+i\omega h_{rx}$  and the prime denotes a derivative with respect to $r$.
The coupling functions $Z,\beta$ depend on $r$ through the scalar field $Z(r)=Z(\varphi(r)),~\beta(r)=\beta(\varphi(r))$.
The axion perturbation equation of $\chi_x$
can be deduced by Eqs.(\ref{eome1}) and (\ref{eome2}) and we do not write it here.

After defining $y=g^2r^4\psi+q a_x$ and removing the variable $\chi_x$, these  perturbation equations  can be further simplified as
\begin{alignat}{1}
&(\frac{f}{\beta g}y^\prime)^\prime+ (\frac{\omega^2 }{r^4\beta gf}-\frac{2m^2}{g^2r^4})y+\frac{2m^2}{g^2r^4}qa_x=0\,,\label{cdeq1}\\
&(Zr^2fa_x^\prime)^\prime+(\frac{\omega^2 Z}{r^2f}-\frac{q^2}{g^2r^4})a_x+\frac{q}{g^2r^4}y=0\,,\label{cdeq2}
\end{alignat}
which are two coupled differential equations of $y,~a_x$.
They take the same form as the differential equations (\ref{gcdeq1}) and (\ref{gcdeq2})  in  Appendix A, with $\frac{\tilde{C}}{C}=2m^2$, $\sqrt{\alpha}=(Z\beta gr^2)^{-1/2}$, and
\begin{alignat}{1}
K=\frac{1}{q}(-g^2r^4((\sqrt{\alpha})^\prime Zr^2f)^\prime+q^2\sqrt{\alpha}-\frac{2m^2}{\sqrt{\alpha}})\,.
\end{alignat}
In general, it is not easy to figure out whether  Eqs.
(\ref{cdeq1}) and (\ref{cdeq2}) can be decoupled. However, if the
quantity $K$ defined above is a constant, then we find it is
completely possible to decouple them,  as discussed in  Appendix
\ref{decouple}.
Here, if the background solution of the electric field satisfies
\begin{alignat}{1}
A_t(r)=\mu-q\sqrt{\alpha}\,,~~or~~\beta=(\frac{1}{\sqrt{\alpha}})^\prime,\label{scd}
\end{alignat}
then $K$  becomes
\begin{alignat}{1}
K
=\frac{1}{q}[r^4g^2(\frac{f}{g})^\prime-q(A_t(r)-\mu)-\frac{2m^2}{\sqrt{\alpha}}].
\end{alignat}
With the use of the equations of motion, one can prove that
$K$ is a constant indeed, which means the perturbation equations
can be decoupled if the condition presented in (\ref{scd}) is
satisfied.
First of all, we point out that the following two
typical solutions in holographic models satisfy this condition.
\begin{itemize}
    \item The first one is simply the standard black brane solution in the EM model with momentum dissipation, which can be
obtained by setting $\varphi=0$, $Z=1$, $\beta=1/2$, and $V=6$
\cite{Andrade:2013gsa,Davison:2015bea}.
    \item The second one is based on the Gubser-Rocha solution in the EMD model in  \cite{Gubser:2009qt}. The momentum
dissipation effect can be introduced by explicitly incorporating
axions into the action
\begin{eqnarray}
S=\int d^4x
\sqrt{-g}\left(R-\frac{e^\varphi}{4}F^2-\sum_{I=x,y}(\partial\phi_I)^2-\frac{3}{2}(\partial\varphi)^2+6\cosh\varphi\right)
\,.
\end{eqnarray}
Setting $\phi_x=mx, \phi_y=my$, and under the ansatz (\ref{anmg}) and
(\ref{bgAt}), this model admits the following solution,\footnote{A
similar result can be obtained for the Gubser-Rocha EMD model in the
framework of massive gravity \cite{Davison:2014}.}
\begin{eqnarray}
\varphi(r)&=&\frac{1}{3}\log g(r)\,,~~f(r)=h(r)g(r)\,,\\
g(r)&=&(1+Q/r)^{3/2}\,,\\
h(r)&=&1-\frac{\mu^2(Q+r_0)^2}{3Q}\frac{1}{(Q+r)^3}-\frac{m^2}{(Q+r)^2}\,,\\
A_t(r)&=&\mu(1-\frac{Q+r_0}{Q+r}) \,,
\end{eqnarray}
where $Q$ is a parameter, and $r_0,~\mu$  represent the position of the horizon
and the chemical potential, respectively. It is easy to check
that this model satisfies the condition (\ref{scd}) and thus its
linearized perturbation equations can be decoupled.
\end{itemize}
Secondly, we should emphasize that the assumption (\ref{scd}) is just a
sufficient condition to ensure the decoupling. Whether it can be
relaxed is still unclear and deserves further investigation.

Before proceeding, we  point out that $K$ being
a conserved quantity implies that it associates with a
scale symmetry. We now elaborate this point as below. It is
more convenient to discuss this problem by using the following
ansatz:
\begin{eqnarray}
  ds^2 &=& -a(r)dt^2+b(r)dr^2+c(r)(dx^2+dy^2)\,, \\
  A &=& A_t(r)dt\,,~~~\phi_x= \phi_x(x)\,,~~~\phi_y= \phi_y(x)\,.
\end{eqnarray}
Substituting them into the action (\ref{AcEMD}), one has
\begin{equation}
S=\int d^4x(\frac{a^2b(c^\prime)^2+2abca^\prime c^\prime+abc^2Z(A_t^\prime)^2}{2(ab)^{3/2}c}-(ab)^{1/2}\beta((\partial_x\phi_x)^2+(\partial_y\phi_y)^2))\,,
\end{equation}
where we have ignored the total derivative terms.
The above action is invariant under the scale transformation
\begin{eqnarray}
c\rightarrow \lambda c\,,~~~ a\rightarrow \lambda^{-2} a\,,~~~ A_t\rightarrow \lambda^{-1} A_t\,,~~~
\phi_x\rightarrow \lambda^{1/2} \phi_x\,,~~~ \phi_y\rightarrow \lambda^{1/2} \phi_y\,.
\end{eqnarray}
Then, similar to the Noether conservation law, we can
obtain the following equation:
\begin{equation}
\partial_r(\frac{1}{\sqrt{ab}}(c^2(\frac{a}{c})^\prime-cZA_t^\prime A_t))-\sqrt{ab}\beta(\partial_x(\phi_x\partial_x\phi_x)+\partial_y(\phi_y\partial_y\phi_y))=0\,.
\end{equation}
For our EMD-axion model, we have $a=1/b=r^2f,~c=r^2g,~\phi_x=m x,~\phi_y=my$, and then the above equation becomes
\begin{equation}
\partial_r(r^4g^2(\frac{f}{g})^\prime-q A_t)-2m^2\beta=0\,,
\end{equation}
where $r^2gZA_t^\prime=q$ has been used. Since we have also introduced the decoupling condition (\ref{scd}),  the above result
becomes a conservation equation,
\begin{equation}
\partial_r(r^4g^2(\frac{f}{g})^\prime-q A_t-\frac{2m^2}{\sqrt{\alpha}})=0\,.
\end{equation}
It means that the fact $K$ is a constant is a
reflection of the scale symmetry of the EMD-axion model.

Since $K$ is a constant under the condition  presented in (\ref{scd}), we may evaluate it at the horizon
$r_0$, which can be rewritten as
\begin{alignat}{1}
K
=\frac{1}{q}[sT+q\mu-\frac{2m^2q}{\mu}]=\frac{1}{q}[\epsilon+P-\frac{2m^2q}{\mu}]
\label{Ke}\end{alignat} where Eq.(\ref{cftq}) and the
relation $\epsilon+P=sT+q\mu$ have been used. Next, following the
discussion at the end of Appendix \ref{decouple} we may decouple
Eqs. (\ref{cdeq1}) and (\ref{cdeq2}) by redefining new variables
as [see Eq. (\ref{decfun})]
\begin{alignat}{1}
v_\pm=\frac{1}{\sqrt{Z\beta r^2g}}(r^4g^2(h_{tx}^\prime+i\omega h_{rx})+qa_x)+\eta_\pm a_x\,,
\label{devar}
\end{alignat}
where the constants $\eta_\pm$ are given by
\begin{alignat}{1}
\eta_\pm=\frac{-K\pm \sqrt{K^2+8m^2}}{2}.\label{dep}
\end{alignat}
Consequently, the decoupled equations of new variables $v_{\pm}$ read as
\begin{eqnarray}
( Zr^2f v^\prime)^\prime+M(\omega)v=0\,,\label{deEq11}
\end{eqnarray}
with
\begin{eqnarray}
M(\omega)=\frac{1}{r^4g^2\sqrt{\alpha}}\left(sT-\frac{2m^2q}{\mu}+\eta q+qA_t(r)\right)+\frac{\omega^2Z}{r^2f}\,,
\end{eqnarray}
where $v_{\pm}$ and $\eta_{\pm}$ have been simply denoted as
$v$ and $\eta$, respectively.

Next, we intend to derive the relation between the currents
sourced by the boundary value of the decoupled variables and the
original momentum $T^{tx}$ and electric current $J^x$, which
are sourced by $h^{(0)}_{tx}$ and $a^{(0)}_x$, respectively.
Before proceeding, it is convenient to introduce new decoupled
variables $s_{\pm}$ as
\begin{eqnarray}
s_+=-\frac{1}{\eta_+}v_+\,,~~~~s_-=\frac{1}{\eta_-}v_-\,,\label{rdss}
\end{eqnarray}
which obey the same differential equation (\ref{deEq11}) as $v_{\pm}$.
In the subsequent section, we  alternately use $v_{\pm}$ and $s_{\pm}$, depending on the convenience of discussion.
We also assume that near the boundary, a generic bulk field perturbation $\Phi(r)$ can be expanded like
\begin{eqnarray}
\Phi(r)=\sum_{n}\frac{\Phi^{(n)}}{r^n}\,.
\end{eqnarray}

Then, with the use of equations of motion (\ref{eome1}) and (\ref{eome2}) as well as expression (\ref{devar}), the pair of new sources $s^{(0)}_\pm$ can be expressed in terms of the original sources as \footnote{Here, we assume
$Z(\varphi)|_{r\rightarrow\infty}=1,~~~~\beta(\varphi)|_{r\rightarrow\infty}=1\,.$}
\begin{eqnarray}
&&
v^{(0)}_\pm=2m^2h_{tx}^{(0)}+\eta_\pm a^{(0)}_x\,,\label{rdss0}
\
\\
&&
s^{(0)}_+=-\frac{1}{\eta_+}v^{(0)}_+\,,~~~~s^{(0)}_-=\frac{1}{\eta_-}v^{(0)}_-\,.\label{rdssb}
\end{eqnarray}
Suppose the corresponding currents sourced by $s^{(0)}_\pm$ are $J_s^\pm$. Since the  transformation of the sources and currents should preserve the form of the perturbed action, namely,
\begin{eqnarray}
T^{tx}h^{(0)}_{tx}+J^x a^{(0)}_x=J_s^+s^{(0)}_++J_s^-s^{(0)}_-\,,\label{iba}
\end{eqnarray}
from the relation of sources in (\ref{rdssb}) one can derive the relation of currents given by
\begin{alignat}{1}
J_s^\pm=\frac{T^{tx}+\eta_\pm J^x}{\eta_--\eta_+}\,.\label{ctr}
\end{alignat}
Now, we have obtained a new pair of sources [Eqs. (\ref{rdss0}) and (\ref{rdssb})] and their corresponding currents [Eq.(\ref{ctr})],
which are the major results from the decoupling process.
So far, they are just a consequence of the special  mathematical structure in the considered  EMD-axion models. Next, we present a brief discussion on their
property and find that this decoupling implies a decomposition of the incoherent and coherent transport.

(1) Because the current $J_s^+$ ($J_s^-$) depends only on $s^{(0)}_+$ ( $s^{(0)}_-$),
there is no overlap between these two currents $J_s^\pm$, i.e., the response function $\chi_{J_s^\pm J_s^\mp}=0$,
which indicates that $J_s^+$ and $J_s^-$ are independent with each other. We thus call them ``independent currents".
As we see in the next sections, this property leads to a decomposition of the conductivity into two individual parts and makes it possible to  obtain an analytic  low
frequency conductivity at small $m$.

(2) For the case of translational invariance, i.e., $m\rightarrow0$,
from Eqs.(\ref{dep}) and (\ref{ctr}), we can see that $J_s^+$ describes the momentum.  Then $J_s^-$ becomes the incoherent current since now it has no overlap with the momentum.
 This coincides with the  result in \cite{Davison:2015taa} and $J^-=J^{inc}$ as
\begin{alignat}{1}
J^{inc}=J_s^-=\frac{sTJ^x-qJ^Q}{\epsilon+P}\,,
\end{alignat}
where all the thermal quantities should take value at $m=0$ and the heat current expression $J^Q=T^{tx}-\mu J^x$ has been used.
It is then clear that the decoupling  represents a decomposition of the incoherent and coherent transport at $m=0$.

In the next section, we begin to
study the conductivity and we  find the decoupling also represents a decomposition of the incoherent and coherent transport for small momentum dissipation.

\section{Conductivities}\label{deG}

In this section, we  first derive the conductivities in terms of the response functions of the independent currents $J_s^\pm$.
Since the modes of sources $s_\pm$ are decoupled,
the matrix of the response function $\chi_s\equiv G_s^R(\omega)-G_s^R(0)$ of currents $J_s^\pm$ is diagonal and we denote $\chi_s=diag(\chi_{s+},\chi_{s-})$.

Under the source transformation (\ref{rdss0}) and (\ref{rdssb}) and the current transformation (\ref{ctr}), the Kubo formula \cite{Hartnoll:2009sz},
\begin{alignat}{1}
\left(
  \begin{array}{c}
    J\\
    J^Q\\
  \end{array}
\right)=\left(
  \begin{array}{cc}
  \sigma_{JJ}\,, & \sigma_{JJ^Q}\\
  \sigma_{J^QJ}\,, &\sigma_{J^QJ^Q}\\
  \end{array}
\right)
\left(
  \begin{array}{c}
    E\\
    -\nabla T/T\\
  \end{array}
\right)
\,,\label{sigmaD}
\end{alignat}
becomes
\begin{alignat}{1}
\left(
  \begin{array}{c}
    J_s^+\\
    J_s^-\\
  \end{array}
\right)=\left(
  \begin{array}{cc}
  \chi_{s+}\,, & 0\\
  0\,, &\chi_{s-}\\
  \end{array}
\right)
\left(
   \begin{array}{c}
    s^{(0)}_+\\
    s^{(0)}_-\\
  \end{array}
\right)
\,.
\end{alignat}
Then, it is straightforward to express the conductivities as
\begin{alignat}{1}
&\sigma_{JJ}=\frac{1}{i\omega}(\chi_{s+} + \chi_{s-})\,,\label{cond1}\\
&\sigma_{JJ^Q}=-\frac{1}{i\omega}((\eta_-+\mu) \chi_{s+}+(\eta_++\mu) \chi_{s-})\,,\\
&\sigma_{J^QJ^Q}=\frac{1}{i\omega}((\eta_-+\mu)^2 \chi_{s+}+(\eta_++\mu)^2 \chi_{s-})\,.\label{cond2}
\end{alignat}
As a consequence,  the  conductivities in Eq. (\ref{sigmaD}) can be calculated through the above equations once
all the response functions $\chi_s$ are known.  Next we  explicitly derive their forms in terms of the perturbation fields following the standard procedure in the context of the linear response theory in the holographic approach.

Usually, we can expand  the action  with respect to some perturbation fields $\Phi_I(k,r)$ as (up to the second order)
\begin{eqnarray}
S^{(2)}[\phi]=\int_{r_B}^{r_H} dr\frac{d^3k}{(2\pi)^3} \Phi^\prime(-k,r)A \Phi^\prime(k,r)+ \Phi^\prime(-k,r)B \Phi(k,r)+\Phi(-k,r)C\Phi(k,r)
\end{eqnarray}
where $kx=-\omega t+\textbf{k}\cdot \textbf{x}$. $\Phi$ represents all $\Phi_I$  and the  matrices  $A(r),~B(r,k),~C(r,k)$ are independent of these perturbation fields.
Then, the perturbed equation of motion of $\Phi(r,k)$ read as \footnote{Taking $A$ as a symmetric matrix has no effect on the results.}
\begin{eqnarray}
(C(r,k)+C^T(r,-k))\Phi=((A+A^T) \Phi^\prime)^\prime - B^T(r,-k) \Phi^\prime+ (B(r,k)\Phi)^\prime\,,\label{PEOM}
\end{eqnarray}
and the retarded Green function as pointed out in \cite{Son:2002} is
\begin{eqnarray}
G^R_{IJ}(k)=-2(r^{-2}A_{IK})^{(0)}\kappa_{KJ}(k)+B_{IJ}^{(0)}(k)\,,\label{Green}
\end{eqnarray}
where the index $0$ again means taking value at the  boundary and $\kappa_{IJ}(k)$ denotes the linear relation $\Phi_I^{(1)}(k)=\kappa_{IJ}(k)\Phi_J^{(0)}(k)$.

Now for $\Phi_I=s=(s_+,s_-)$ with $\textbf{k}=0$,  the equations in (\ref{PEOM}) should be decoupled and become
\begin{eqnarray}
(A s^\prime)^\prime +\frac{1}{2}(B(r,\omega)- B(r,-\omega))s^\prime+\frac{1}{2}(B^\prime(r,\omega)-C(r,\omega)-C(r,-\omega))s=0\,.
\end{eqnarray}
Comparing the above equation with Eq.(\ref{deEq11}) (which is also the differential equation of $s$ ),
one has $B(r,\omega)=0$ and $A$ is diagonal,  whose elements, denoted as $diag(A_{++},A_{--})$, are proportional to $Zr^2f$.  We can then  choose
\begin{eqnarray}
&&(r^{-2}A_{++})^{(0)}=(r^{-2}A_{--})^{(0)}=-\frac{\Lambda_{\pm}}{2}\,,
\end{eqnarray}
where $\Lambda_\pm$ are arbitrary constants.
Therefore, the  response function $\chi_s$  can be derived through Eq.(\ref{Green}),
\begin{eqnarray}
\chi_{s\pm}(\omega)=\Lambda_\pm\left(\frac{s_\pm^{(1)}(\omega)}{s_\pm^{(0)}(\omega)}-\frac{s_\pm^{(1)}(0)}{s_\pm^{(0)}(0)}\right)\,.
\end{eqnarray}

Finally, we can obtain the conductivities in terms of the perturbation fields through Eqs.(\ref{cond1})-(\ref{cond2}) as
\begin{alignat}{1}
&\sigma_{JJ}=\frac{1}{i\omega}(\Lambda_+\Theta_++\Lambda_-\Theta_-)\label{cond11}\\
&\sigma_{JJ^Q}=-\frac{1}{i\omega}((\eta_-+\mu)\Lambda_+\Theta_++(\eta_++\mu)\Lambda_-\Theta_-)\,,\\
&\sigma_{J^QJ^Q}=\frac{1}{i\omega}((\eta_-+\mu)^2\Lambda_+\Theta_++(\eta_++\mu)^2\Lambda_-\Theta_-)\label{cond22}\,.
\end{alignat}
where
\begin{eqnarray}
\Theta_\pm\equiv\frac{s_\pm^{(1)}(\omega)}{s_\pm^{(0)}(\omega)}-\frac{s_\pm^{(1)}(0)}{s_\pm^{(0)}(0)}\,.
\end{eqnarray}
Note that the expression of $\Theta_\pm$ is invariant under the change of $s\rightarrow v$.
Up to the current stage, we have found that the frequency dependence of conductivities can be determined once the quantities $\Theta_\pm$ and $\Lambda_\pm$ are known.
Next we  demonstrate that $\Theta_\pm$ can be obtained by solving the decoupled equation (\ref{deEq11}),
while $\Lambda_\pm$ can be determined by the DC conductivity and the incoherent conductivity.

\section{The low frequency behaviors of conductivities}\label{Lfc}

In this section we  study the property of conductivities in the low frequency limit.
The decoupling results in Sec.\ref{EMDa} are a key step to obtain the analytic low frequency  conductivities, since
they decompose the conductivity into two individual contributions coming from $J^\pm$,  respectively. Furthermore, we
 find that one part of the conductivity exhibits a Drude behavior corresponding to a coherent transport while
the other part is dominant by the incoherent transport. Then, we  not only give an analytic expression of
the low frequency conductivities , but also achieve a decomposition of the incoherent and coherent transport.

To derive the low frequency conductivities, according to the results in Sec.\ref{deG}
 , we  derive the expression of $\Theta_\pm$ by approximately solving decoupled equations (\ref{deEq11}) with low frequency expansion.
For this purpose, it is convenient to write $v$ as
\begin{alignat}{1}
v=f^{-i\omega/(4\pi T)}v_0(r)F(r)\,,\label{vex}
\end{alignat}
where $v_0$ is a zero frequency solution to  Eq.(\ref{deEq11}) with $\omega=0$, which can be chosen as
\begin{alignat}{1}
v_0(r)\equiv1+\frac{q}{\eta}\sqrt{\alpha}\,,
\end{alignat}
and $F(r)$ is regular at the horizon which can be expanded as
\begin{alignat}{1}
F(r)=F_0+F_1(r)\omega+\mathcal{O}(\omega^2)\,.
\end{alignat}
For simplicity, here  we still use one character to denote the two solutions  of  Eq.(\ref{deEq11}). Since the two equations are only different with parameter $\eta_{\pm}$, we can write, for instance, $F=(F_+(\eta_+),F_-(\eta_-))$ and
so on.
Without loss of generality, we can choose $F_0=1$.
Then $\Theta=(\Theta_+,\Theta_-)$ can be expressed as
\begin{alignat}{1}
\Theta=\frac{v^{(1)}(\omega)}{v^{(0)}(\omega)}-\frac{v^{(1)}(0)}{v^{(0)}(0)}=\frac{F_1^{(1)}\omega+\mathcal{O}(\omega^2)}{1+F_1^{(0)}\omega+\mathcal{O}(\omega^2)}\label{ff}
\,.
\end{alignat}

By substituting the expression in (\ref{vex}) into Eqs.(\ref{deEq11}),
we can derive the following equation of $F_1$\footnote{ We have assumed $r^2f^\prime=0$ at $r=\infty$.}
\begin{alignat}{1}
F_1^\prime(r)=i\frac{Zr^2v_0^2f^\prime-4\pi T(Zv_0^2)|_{r_+}}{4\pi TZr^2fv_0^2 }\,,\label{fdeEq11}
\end{alignat}
whose solution provides results to $F_1^{(0)},F_1^{(1)}$ and thus determines the low frequency conductivity.
$F_1^{(1)}$ can be obtained directly from Eq.(\ref{fdeEq11}) as
\begin{alignat}{1}
F_1^{(1)}=i(Zv_0^2)|_{r_+}=iZ(r_+)\left(1+\frac{\mu}{\eta_\pm}\right)^2\,,\label{Fp}
\end{alignat}
which determines the DC conductivity. $F_1^{(0)}$, which is usually hard to solve,  can be obtained at small $m$.

We first discuss the DC conductivity, which reads as [with the use of Eq.(\ref{cond11})]
\begin{alignat}{1}
\sigma_{DC}=\Lambda_+Z(r_+)\left(1+\frac{\mu}{\eta_+}\right)^2+\Lambda_-Z(r_+)\left(1+\frac{\mu}{\eta_-}\right)^2
\,.\label{DC1}
\end{alignat}
Obviously, the above expression is decomposed into two
contributions from the independent currents $J^{\pm}$. When $m=0$, the first part from the momentum
contribution is divergent due to the translational
invariance, while the second part provides the incoherent
conductivity which should be the same as   Eq.(\ref{sigmainc}),
\begin{alignat}{1}
\sigma_{inc}=Z(r_+)\Big(\frac{sT}{\epsilon+P}\Big)^2\Big|_{m=0}\equiv\sigma_Q\,.
\end{alignat}
Then the
coefficient $\Lambda_-$ can be determined as an expansion form of
slow momentum  $\Lambda_-=1+\mathcal{O}(m^2)$. Now we can further
determine $\Lambda_+$ by comparing the expression in
(\ref{DC1}) with the DC conductivity result calculated via the
horizon data \cite{Donos:20146,Donos:2014cya},
\footnote{We can also refer to \cite{Vegh:20131,Donos:201311,Blake:20138,Donos:201406,DC1,DC2} for the analytical calculation on DC conductivity of holographic systems with momentum dissipation.}
\begin{alignat}{1}
\sigma_{DC}=Z(r_+)\Big(1+\frac{\mu^2}{2m^2}\Big)\,.\label{DC2}
\end{alignat}
In the case of slow momentum dissipation, we can simply rewrite
Eq.(\ref{DC1}) as
\begin{alignat}{1}
\Lambda_+Z(r_+)\Big(1+\frac{\mu}{\eta_+}\Big)^2=\sigma_{DC}-\sigma_Q+\mathcal{O}(m^2)\,.\label{pG1}
\end{alignat}

Next, we study the low-frequency behavior of conductivities with
slow momentum dissipation (small $m$).  Explicitly, we  treat $\omega$ and $m^2$ as the same
order. Usually it is hard to
obtain an analytical solution of $F_1^{(0)}(r)$ from Eq.(\ref{fdeEq11}). Nevertheless,
what we are mainly concerned with is whether there exists the
electric current dissipation, which can be signaled by the
appearance of a pole in the expression of conductivity. Observing the expression in Eq.(\ref{ff}), we find only
when $F_1^{(0)}$ becomes divergent as $m\rightarrow0$, the
conductivity may have a pole structure.

For $\eta=\eta_-$, since $F^\prime_1(r) $ is regular as
$m\rightarrow0$, and so does $F_1^{(0)}$, there is no pole
structure in  $\Theta_-$ and then we can write
\begin{alignat}{1}
\Theta_-=\sigma_Q+\mathcal{O}(m^2,\omega)\,,
\end{alignat}
without losing the interesting properties. This implies that $J^-$ can be an
incoherent current at small $m$, since its contribution to the conductivity is dominant by the incoherent conductivity

For $\eta=\eta_+$,  $F_1^{(0)}$ is divergent  as
$m\rightarrow0$. Thus the pole structure totally comes from the contribution of  $J^+$, which implies that $J^+$  is a coherent current, describing the dissipation process.
To solve $F_1^{(0)}$ for $\eta=\eta_+$,
it is convenient to introduce a new variable $\psi(r)$
\begin{alignat}{1}
\psi(r)\equiv F_1(r)-i\frac{1}{4\pi T}ln f(r)\,,
\end{alignat}
whose asymptotic behavior is like that of $F_1$
\begin{alignat}{1}
\psi^{(0)}=F_1^{(0)}\,,~~~~\psi^{(1)}=F_1^{(1)}\,.
\end{alignat}
Then,  to obtain $F_1^{(0)}$, we need only deal with the equation of $\psi(r)$, which is given from Eq.(\ref{fdeEq11}) as
\begin{alignat}{1}
\psi^\prime(r)=-i\frac{(Zv_0^2)|_{r_+}}{Zr^2fv_0^2 }\,.\label{fdeEq111}
\end{alignat}
Integrating the above equation from the horizon to the boundary,
we have
\begin{eqnarray}
&&
\psi^{(0)}
=-i(Z_+(\eta_++\mu)^2)\int_{r_+}^\infty\frac{dr}{Zr^2f(\eta_++q\sqrt{\alpha})^2 }\,,
\nonumber
\\
&&
=-i(Z_+(\eta_++\mu)^2)\Big(\frac{g}{qf(\eta_++q\sqrt{\alpha})}\Big|_{r_+}^\infty-\int_{r_+}^\infty(\frac{g}{f})^\prime\frac{dr}{q\Big(\eta_++q\sqrt{\alpha}\Big) }\Big)\,,
\nonumber
\\
&&
=-\frac{i(Z_+(\eta_++\mu)^2)}{q\eta_+}+\mathcal{O}(1)\,,\label{O1}
\nonumber
\\
&&
\equiv-i\Gamma^{-1}+\mathcal{O}(1)\,.
\end{eqnarray}
Note that the divergence at $r_+$ is not a genuine singular and
would not affect our discussion since the in-falling condition
guarantees the regularity of $F_1$. As a result, we
have
\begin{alignat}{1}
\Theta_+=\frac{F_1^{(1)}\omega+\mathcal{O}(\omega^2)}{1-i\omega\Gamma^{-1}+\mathcal{O}(1)\omega+\mathcal{O}(\omega^2)}
\,,
\end{alignat}
with
\begin{alignat}{1}
\Gamma&=\frac{q\eta_+}{Z_+(\eta_++\mu)^2}=\frac{2m^2s\beta}{4\pi(\epsilon+P)}+\mathcal{O}(m^4)\,.\label{relax}
\end{alignat}
Therefore, $\Theta_+$  exhibits a Drude behavior at low frequency regime
with the dissipation rate $\Gamma\sim m^2$ and thus is the coherent contribution to the conductivities.  Note that since we have neglected the
subleading term $\mathcal{O}(1)$ in (\ref{O1}), which is  the
same order as $\mathcal{O}(m^4)$ in   (\ref{relax}), the above
expression is only viable at $\mathcal{O}(m^2)$.

Finally, combining the incoherent and coherent contribution and  using Eq. (\ref{pG1}) and the formulas in Eqs.
(\ref{cond11})-(\ref{cond22}) we obtain the low frequency conductivity at small momentum dissipation
\begin{alignat}{1}
&\sigma_{JJ}=\frac{\sigma_{DC}-\sigma_Q+\mathcal{O}(\Gamma,,\omega)}{1-i\omega/\Gamma}+\sigma_Q+\mathcal{O}(\Gamma,\omega)\\
&\sigma_{JJ^Q}=\frac{\frac{sT}{q}(\sigma_{DC}-\sigma_Q)+\mathcal{O}(\Gamma,,\omega)}{1-i\omega/\Gamma}-\mu\sigma_Q+\mathcal{O}(\Gamma,\omega)\,,\\
&\sigma_{J^QJ^Q}=\frac{\frac{s^2T^2}{q^2}(\sigma_{DC}-\sigma_Q)+\mathcal{O}(\Gamma,,\omega)}{1-i\omega/\Gamma}+\mu^2\sigma_Q+\mathcal{O}(\Gamma,\omega)\,.
\end{alignat}
The above results can also be rewritten as
\begin{alignat}{1}
&\sigma_{JJ}=\frac{\frac{q^2}{\epsilon+P}+\mathcal{O}(\Gamma,\omega\Gamma,\omega^2)}{\Gamma-i\omega}+\sigma_Q+\mathcal{O}(\Gamma,\omega)\\
&\sigma_{JJ^Q}=\frac{\frac{qsT}{\epsilon+P}+\mathcal{O}(\Gamma,\omega\Gamma,\omega^2)}{\Gamma-i\omega}-\mu\sigma_Q+\mathcal{O}(\Gamma,\omega)\,,\\
&\sigma_{J^QJ^Q}=\frac{\frac{s^2T^2}{\epsilon+P}+\mathcal{O}(\Gamma,\omega\Gamma,\omega^2)}{\Gamma-i\omega}+\mu^2\sigma_Q+\mathcal{O}(\Gamma,\omega)\,.
\end{alignat}
Now, we see clearly that the decoupling process in Sec.\ref{EMDa} leads to a coherent and incoherent decomposition
of the conductivities at small momentum dissipation. While the current $J^+$ provides a  coherent Drude contribution,  $J^-$
responds to the incoherent contribution. This means that at small $m$, the transport
comes from two individual contributions, the dissipation and diffusion process, captured by the decoupling currents $J^\pm$, respectively.

We also give the analytic low frequency conductivities at small momentum dissipation for a large class of EMD-axion thories
that coincide with the results derived from
the hydrodynamic memory matrix technique up to the leading order. It
is the first time to obtain the analytic low frequency conductivities, which is valid for so many theories in holography. It also confirms that a hydrodynamical description of dual field in holography is usually suitable.

\section{conclusions and discussions}\label{con}

In this paper we have investigated the incoherence of conductivity
in the holographic framework  explicitly in terms of  EMD-axion theory. The key ingredient in this
analysis is the decoupling of the linearized perturbation
equations. Based on the decoupled equations, we have introduced a
new pair of independent currents $J_s^\pm$. When momentum is
conserved, they go back to the momentum and the incoherent current
$J^{inc}$ respectively, which coincides with the discussion about
incoherence in \cite{Hartnoll:2014lpa}. For slow momentum
dissipation, we have derived the analytic expression for
conductivity in the low frequency regime, which can be viewed as a
generalization of the results presented in \cite{Davison:2015bea}.
We have demonstrated that it can also be divided into two
parts in the EMD-axion model as that in the Einstein-Maxwell model \cite{Davison:2015bea}, including the coherent contribution from $J^+$ and the
incoherent one from $J^-$. Our holographic argument here also
confirms the general results derived in the context of
hydrodynamics up to the leading order of low frequency
expansion.

Our work has made progress on the understanding of the incoherent
part of conductivity in a more general holographic circumstance.
It would be quite possible to push our investigation forward along the
following directions in the future.
\begin{itemize}
\item In the fluid/gravity approach, the same results of low frequency
conductivity as in \cite{Davison:2015bea} have been obtained for
the EM-axion theory in \cite{Blake:20155}. It is very worth
extending this to EMD-axion models and then making a comparison with
our results in the current paper, which might be helpful for us to
understand the physical implication of the decoupling condition as
proposed in Eq.(\ref{scd}).

\item Recently, from the memory matrix theory \cite{Lucas:2015pxa}
as well as  the fluid/gravity approach \cite{Blake:20157}, the low
frequency behavior of the conductivity in magnetotransport has
been obtained. It is also worth  exploring this subject from the
incoherent and coherent point of view.

\item As we know, the assumption (\ref{scd}) is applicable
to the EM-axion model and Gubser-Rocha EMD-axion model. It would be
interesting to figure out if we could find more background
solutions subject to this condition in a wide range of holographic
models such as hyperscaling-violating models.

\end{itemize}

\begin{acknowledgments}
We are grateful to Peng Liu for helpful discussion on the
incoherence of conductivity. This work is supported by the Natural
Science Foundation of China under Grants No.11275208,  No.11305018, and
No.11178002. Y.L. also acknowledges the support from Jiangxi young
scientists (JingGang Star) program and 555 talent project of
Jiangxi Province. J. P. W. is also supported by the Program for
Liaoning Excellent Talents in University (Grant No. LJQ2014123).

\end{acknowledgments}

\begin{appendix}

\section{Useful tool for decoupling}\label{decouple}

In this appendix, we  discuss the decoupling condition
of two coupled differential equations with the following
general form,
\begin{alignat}{1}
\label{gcdeq1}
&(\alpha A y^\prime)^\prime+\tilde{B}y+\tilde{C}a_x=0\,,\\
\label{gcdeq2}
&(Aa_x^\prime)^\prime+Ba_x+Cy=0\,,
\end{alignat}
where equation variables $y,~a_x$ are coupled to each other,
and $A,~\alpha,~B,~C,~\tilde{B},~\tilde{C}$ are all functions of
$r$. Introducing a new variable $u=y/k(r)$ with $k=
\frac{1}{\sqrt{\alpha}}$, the above equations become
\begin{alignat}{1}
\label{gcdeq11}
&( A u^\prime)^\prime+(\frac{\tilde{B}}{\alpha}-\frac{1}{\sqrt{\alpha}}(\frac{\alpha^\prime}{2\sqrt{\alpha}} A )^\prime)u+\frac{\tilde{C}}{\sqrt{\alpha}}a_x=0\,,\\
\label{gcdeq12}
&(Aa_x^\prime)^\prime+Ba_x+\frac{C}{\sqrt{\alpha}}u=0\,.
\end{alignat}
It can be easily found that if there is a constant $\eta$
satisfying
\begin{alignat}{1}
\frac{\tilde{B}}{\alpha}-\frac{1}{\sqrt{\alpha}}(\frac{\alpha^\prime}{2\sqrt{\alpha}} A )^\prime+\eta\frac{C}{\sqrt{\alpha}}=B+\frac{\tilde{C}}{\sqrt{\alpha}}\frac{1}{\eta}\,,
\end{alignat}
then, multiplying Eq. (\ref{gcdeq12}) with the constant $\eta$ and
combining it with Eq. (\ref{gcdeq11}), we can obtain the
following decoupled equations,
\begin{alignat}{1}
( A v^\prime)^\prime+(\frac{\tilde{B}}{\alpha}-\frac{1}{\sqrt{\alpha}}((\sqrt{\alpha})^\prime A )^\prime+\eta\frac{C}{\sqrt{\alpha}})v=0\,,\label{deEq}
\end{alignat}
with the decoupled variables defined as
\begin{alignat}{1}
v=u+\eta a_x=\sqrt{\alpha}y+\eta a_x\,.\label{decfun}
\end{alignat}
The constant $\eta$ is determined by the following equation:
\begin{alignat}{1}
&\eta^2+K\eta-\frac{\tilde{C}}{C}=0\,,\label{dec}\\
&K\equiv\frac{\tilde{B}}{C\sqrt{\alpha}}-\frac{1}{C}((\sqrt{\alpha})^\prime A )^\prime-\frac{B\sqrt{\alpha}}{C}\,.
\end{alignat}
That is to say, if above Eq. (\ref{dec}) has a constant solution,
the perturbation equations (\ref{gcdeq1} and \ref{gcdeq2}) can be decoupled
into Eq. (\ref{deEq}). A simple example of Eq. (\ref{dec}) with a
constant $\eta$ solution is that the coefficients
$\frac{\tilde{C}}{C}$ and
$K=\frac{\tilde{B}}{C\sqrt{\alpha}}-\frac{1}{C}((\sqrt{\alpha})^\prime
A )^\prime-\frac{B\sqrt{\alpha}}{C}$ are both constants. Note that
the decoupling conditions of the coupled equations
(\ref{gcdeq1}) and (\ref{gcdeq2})  we discussed here are only the
sufficient conditions, not necessary conditions.

\end{appendix}

\end{document}